\documentclass[
    ,final            % use final for the camera ready runs
  ]
  {aipproc}

\layoutstyle{6x9}

\begin{document}

\title{Diffractive jets production in pp-collisions}


\author{D.Yu. Ivanov}{
  address={Sobolev Institute of
Mathematics and Novosibirsk State University, 630090 Novosibirsk,
Russia} }

\author{V.M. Braun}{
  address={Institut f\"ur Theoretische Physik, Universit\"at Regensburg,
D-93040 Regensburg, Germany} }

\author{A.
Sch\"afer} {
  address={Institut f\"ur Theoretische Physik, Universit\"at Regensburg,
D-93040 Regensburg, Germany}
%  ,altaddress={<author1 address>} % additional visiting address
}

\begin{abstract}
We consider the exclusive diffractive dissociation of a proton into
three jets with large transverse momenta in the double-logarithmic
approximation of perturbative QCD. This process is sensitive to the
proton unintegrated gluon distribution at small x and to the proton
light-cone distribution amplitudes. According to our estimates, an
observation of such processes in the early runs at LHC is feasible
for jet transverse momenta of the order of $5$~GeV.

\end{abstract}

\maketitle

%%%%%%%%%%%%%%%%%%%%%%%%%%%%%%%%%%%%%%%%%%%%
%% MAINMATTER
%%%%%%%%%%%%%%%%%%%%%%%%%%%%%%%%%%%%%%%%%%%%

{\bf 1.}~~ We explore the possibility to observe hard exclusive
diffractive dissociation of a proton into three hard jets in
proton-proton collisions
\begin{equation}
p(p_1)+p(p_2)\to jet(q_1)+jet(q_2)+jet(q_3)+ p(p_2^\prime)\, .
\label{process}
\end{equation}
In this process one proton stays intact and the other one
dissociates into a system of three hard jets separated by a large
rapidity gap from the recoil proton, see Fig.~1.

Note that  we are interested in {\it exclusive}\  three--jet
production which constitutes a small fraction of the inclusive
single diffraction cross section. The exclusive and inclusive
mechanisms have  different final state topologies and can be
distinguished experimentally. A characteristic quantity is e.g. the
ratio $R_{jets}$ of the three-jet mass to the total invariant mass
of the system produced in the diffractive interaction. Exclusive
production corresponds to  the region where $R_{jets}$ is close to
unity. This strategy was used recently at the Tevatron
\cite{Aaltonen:2007hs} where  central exclusive dijet production,
$p\bar p\to p +jet+jet+\bar p$, in double--Pomeron collisions was
measured for the first time.

Exclusive dijet production in the central region has much in common
with the exclusive Higgs boson production process, $p\bar p\to p
+H+\bar p$. In \cite{Khoze:2008cx} it was argued that studies of
exclusive dijet production and other diffractive processes at the
early data runs of the LHC can provide valuable checks of the
different components of the formalism. Indeed, this was the main
motivation for Tevatron experiment. The exclusive 3-jets production
in single diffraction (\ref{process}) offers another interesting
example since factorization of hard and soft interactions in this
case is less complicated. In particular,  the fluctuation of a
proton projectile into  a state with small transverse size, which is
the underlining mechanism for  (\ref{process}), suppresses secondary
soft interactions that may fill the rapidity gap. Thus one can get
an access to the gluon distribution at small $x$ in a cleaner
environment, having no problems with gap survival probability and
factorization breaking that introduce major conceptual theoretical
uncertainties in the calculations of diffractive Higgs production.

Our approach to exclusive three-jet production derives from
experience with coherent pion diffraction dissociation into a pair
of jets with large transverse momenta which was measured by the E791
collaboration \cite{E791a,E791b}. The qualitative features of the
E791 data have confirmed
 some earlier theoretical predictions \cite{KDR80,BBGG81,FMS93}:
a strong A-dependence which is  a signature for color transparency,
and a $\sim 1/q_\perp^8$ dependence on the jet transverse momentum.
These features suggest that the relevant transverse size of the pion
$r_\perp$ remains small, of the order of the inverse transverse
momenta of the jets $r_\perp  \sim 1/q_\perp$.

We have shown \cite{Braun:2001ih,Braun:2002wu} that collinear
factorization is violated in dijet production due to pinching of
singularities between soft gluon (and quark) interactions in the
initial and final state. However, the nonfactorizable contribution
is suppressed compared to the leading contribution by a logarithm of
energy so that
 in  the double logarithmic approximation  $\ln q_\perp^2 \ln s/q_\perp^2$
 collinear factorization is restored.
Moreover, to this accuracy  hard gluon exchange can be ``hidden'' in
the  unintegrated gluon distribution ${\cal F}(x,q_\perp)$. Thus, in
the true diffraction limit, for very large energies, hard exclusive
dijet production can be considered as a probe of the hard component
of the pomeron. The same interpretation  was suggested  earlier in
\cite{NSS99} within the $k_t$ factorization framework. The double
logarithmic approximation turns out to be insufficient for the
energy range of the E791 experiment, but might be adequate for the
LHC. Here we present an estimate for the cross section for the
reaction (\ref{process}) based on the generalization of these ideas.

\begin{figure}
  \includegraphics[height=.2\textheight]{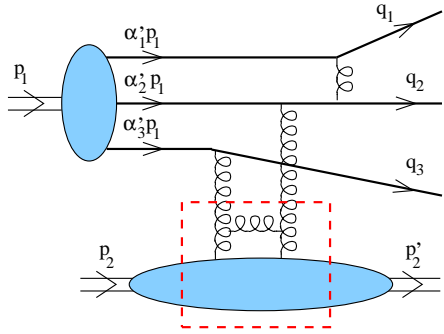}
  \caption{Proton
dissociation into three jets. The unintegrated gluon distribution
includes the hard gluon exchange as indicated by the dashed square.}
\end{figure}

\vskip0.2cm

{\bf 2.}~ At leading order the jets are formed by the three valence
quarks of the proton, see Fig.~1.
 We require that all three jets
have large transverse momenta which requires at least two hard gluon
exchanges. One of them can be effectively included in the
high-momentum component of the unintegrated gluon density (the
bottom blob) as indicated schematically by the dashed square,
 but the second one has to be added explicitly since the  hard  pomeron only couples to two
of the three quarks of the proton.

Our notation for the momenta is explained in Fig.~1. We neglect
power corrections in  transverse momenta of the jets and also proton
and jet masses so that
%\begin{equation}
$p_1^2=p_2^2=p_2^{\prime 2}=q_1^2=q_2^2=q_3^2=0.$
%\end{equation}
The jet momenta are decomposed in terms of momenta of the initial
particles
\begin{eqnarray}
q_k=\alpha_k p_1+\beta_k p_2+q_{k\perp}\,, \qquad k=1,2,3 \, .
%\nonumber \\
%q_2=\alpha_2 p_1+\beta_2 p_2+q_{2\perp}\, , \nonumber \\
%q_3=\alpha_3 p_1+\beta_3 p_2+q_{3\perp}\, .
\end{eqnarray}
The  three--jet invariant mass is given by
\begin{equation}
\label{zeta} M^2=(q_1+q_2+q_3)^2=\frac{\vec q_{1\perp}^{\,\,
2}}{\alpha_1} +  \frac{\vec q_{2\perp}^{\,\, 2}}{\alpha_2}+
\frac{\vec q_{3\perp}^{\,\, 2}}{\alpha_3}\, ,\qquad
\zeta=\frac{M^2}{s} = \beta_1+\beta_2+\beta_3\,.
\end{equation}
where $s=(p_1+p_2)^2=2p_1\cdot p_2$ is the invariant energy.
Assuming that the  relevant  jet transverse momenta are of the order
of 5~GeV, the typical values of the $\zeta$ variable at LHC are in
the range $\zeta\sim 10^{-6}\div 10^{-5}$.

The relevant Feynman diagrams can be divided into three groups which
differ by the attachments of the $t$-channel gluons to the quark
lines. Accordingly, we have three different contributions to the
amplitude (for the details see \cite{Braun:2008ax}):
\begin{eqnarray}
\label{ampl}
 \mathcal{M} & = & -i\,   2^7 \pi^5\, s \, \alpha_s^2
\left[\frac{e^{ijk}\left(\frac{1+N}{N}\right)^2}{4N!(N^2-1)}\right]\,
\int D\alpha^\prime\, \times \left(
\mathcal{L}_{1}\frac{\delta(\alpha_1-\alpha_1^\prime)}{q_{1\perp}^4}{\mathcal
F}(\zeta,q_{1\perp})\right.
 \nonumber \\
&&\left. \hspace*{-1cm}+
\mathcal{L}_{2}\frac{\delta(\alpha_2-\alpha_2^\prime)}{q_{2\perp}^4}{\mathcal
F}(\zeta,q_{2\perp})+
\mathcal{L}_{3}\frac{\delta(\alpha_3-\alpha_3^\prime)}{q_{3\perp}^4}{\mathcal
F}(\zeta,q_{3\perp})\right) \, ,
\end{eqnarray}
where $\int D\alpha' = \int_0^1 d\alpha'_1 d\alpha'_2 d\alpha'_3
\delta(1-\sum \alpha'_i)$ corresponds to the integration over the
quark momentum fractions in the incident proton, $e^{ijk}$ describes
the color state of the final quarks, $N=3$ is the number of colors.
The dimensionless quantities $\mathcal{L}_i$ are expressed in terms
of  the leading-twist light-cone nucleon distribution amplitudes.

\vskip0.2cm

{\bf 3.}~The differential cross section can be written as
\begin{equation}
d\sigma =\frac{|\mathcal{M}|^2}{2^5(2\pi)^8 s^2} \frac{d\alpha_1
d\alpha_2 d\alpha_3
\delta(1-\alpha_1-\alpha_2-\alpha_3)}{\alpha_1\alpha_2\alpha_3}
d^2\vec q_{1}d^2\vec q_{2} dt d\phi_t
\end{equation}
where $t=(p_2-p_2^\prime)^2$ is the Mandelstam $t$ variable of the
$pp$ scattering and $\phi_t$ is the azimuthal angle of the final
state proton. In our kinematics, for large transverse momenta of the
jets and small $t$, one can neglect effects of azimuthal
correlations between the jets and the final proton. Hence  $d\phi_t$
integration is trivial and gives a factor $2\pi$. For the $t$
dependence we assume a simple exponential form, $d\sigma/dt \sim
e^{b t}$, and use $b\sim 4\div 5 \, \rm{GeV}^2$ for the slope
parameter which is a typical value which describes HERA data for
hard exclusive processes: DVCS and vector meson electroproduction at
large $Q^2$. Thus, the integration over the proton recoil  variables
gives a factor $\int dt d\phi_t \to \frac{2\pi}{b}$.

Since our calculation is only done to double logarithmic accuracy,
 we use the simplest model for the unintegrated gluon distribution as given by the
logarithmic derivative of the usual gluon parton distribution $x g
(x,Q^2)$
\begin{equation}
\label{unintegrated} {\mathcal F}(x,q_{\perp}^2)={\partial \over
\partial \ln q_\perp^2}\, x\, g(x,q_\perp^2) \,.
\end{equation}
In our numerical estimates we use the CTEQ6L leading-order gluon
distribution \cite{Pumplin:2002vw}.

The integration over the phase space of the three jets was done
numerically, restricting the
longitudinal momentum fractions to the region
\begin{equation}
0.1 \leq \alpha_1,\alpha_2,\alpha_3 \leq 0.8
\end{equation}
and requiring that the transverse momentum of {\it each} jet is
larger than a given value $q_0 = q_{\perp, {\rm min }}$. For $q_0 =
5$~GeV we obtain for the integrated three-jet cross section at the
LHC energies
\begin{equation}
 \sigma^{\rm LHC}_{3-jets} = 4\,\mbox{pb}\,\cdot \left(\frac{f_N(q_0)}{4.7\cdot 10^{-3}\mbox{\small GeV}^2}\right)^2
 \left(\frac{\alpha_s(q_0)}{0.21}\right)^4 \left(\frac{5 \,\mbox{\small GeV}}{q_0}\right)^9.
\end{equation}
Assuming the integrated luminosity for the first LHC runs in the
range $100$~pb$^{-1}$ to  $1$~fb$^{-1}$ an observation of this
process at LHC seems to be feasible. Note that the effective power
$\sigma \sim1/q^9_0$ (fitted in the $q_0=3\div 8$~GeV range) is
somewhat stronger than the naive power counting prediction
$\sigma\sim 1/q_0^8$. This effect is due to the strong $\zeta$
dependence of the unintegrated gluon distribution: larger values of
$q_0$ imply larger invariant masses $M^2$ of the three-jet system
(\ref{zeta}) and consequently  larger $\zeta=M^2/s$. The sizeable
cross section for $q_0=5$~GeV is in fact an implication of the
expected rise of the LO gluon distribution more than two times as
$\zeta$ is decreasing by roughly a factor of $50$ when going from
Tevatron to LHC.
The existing parameterizations of the LO gluon distribution at
$\zeta\sim 10^{-6}$ differ from each other by $\sim 30\%$. The
unintegrated gluon distribution (\ref{unintegrated}) enters as a
square in the prediction for the cross section, therefore, the study
of exclusive three-jet events at LHC may provide a valuable
constraint for the gluon distribution at small momentum fractions.

A comparison of the three-jet exclusive production at LHC and the
Tevatron can be especially illuminating in this respect since other
uncertainties do not have significant impact on  the energy
dependence. For Tevatron kinematics, assuming the value $q_{\perp
\rm{min}}= 3$~GeV, our estimate for the cross section (fitted in the
range $q_0=2\div 4.5$~GeV) is
\begin{equation}
\label{tevatron}
 \sigma^{\rm Tevatron}_{3-jets} = 50\,\mbox{pb}\,\cdot
 \left(\frac{f_N(q_0)}{4.7\cdot 10^{-3}\mbox{\small GeV}^2}\right)^2
 \left(\frac{\alpha_s(q_0)}{0.255}\right)^4 \left(\frac{3\,\mbox{\small GeV}}{q_0}\right)^9.
\end{equation}

\begin{theacknowledgments}
D.~I.\ was supported in part by the grants RFBR-08-02-00334-a and
NSh-1027.2008.2.
\end{theacknowledgments}

\end{document}